\documentclass{elsart1p}
\usepackage{graphicx}
\usepackage{amssymb}

\begin{document}
\begin{frontmatter}
%
%
%
\title{Status of CP Violation}
%
%
\author{Nita Sinha}
\address{The Institute of Mathematical Sciences, Chennai, India}
\begin{abstract}
  The Standard Model parametrization of CP violation is described.
  Tests of this parametrization using the observed heavy flavour
  decays and implications for New physics are discussed.
\end{abstract}
\begin{keyword}
CP violation, B mesons, New Physics
%
\PACS 11.30Er, 11.30Hv, 12.15Hh, 13.20He
\end{keyword}
\end{frontmatter}
%
\section{A bit of history}
\label{}
With the announcement of the 2008 Nobel prize in Physics, there has
been a lot of excitement among the community working on CP violation
(CPV). In 1972, Kobayashi and Maskawa had proposed that CPV could be
incorporated as a single phase in the three generation quark mixing
matrix. With only three quarks $(u,d,s)$ known at that time, it had
been a bold step. CPV had been discovered in 1964, as a tiny effect in
the decays of $K$ mesons by Cronin and Fitch. CPV has an important
role to play in the matter-antimatter asymmetry of the universe.
It is believed that the Universe was born with equal amounts of matter
and antimatter, but since we only see matter around us, it implies
that matter and antimatter behave differently. In 1967, Sakharov gave
the famous three conditions for generating a baryon number asymmetry
in the universe: 
\begin{itemize}
\item [$\bullet$] Baryon number violation, to allow the antibaryons to
  disappear while baryons survive.
\item [$\bullet$] CPV, since the decay rates of baryons and
  antibaryons must differ.
\item [$\bullet$] Departure from thermal equilibrium, to ensure that the
  created asymmetry is not washed away.
\end{itemize}

Ten years after Sakharov laid down the above stated conditions, the
bound state of the $b$ quark, the Upsilon was observed. In 1984, Bigi
and Sanda showed that in the Kobayashi and Maskawa (KM) picture, the
CP violating effects would be observable in the $B$ system if, the $B$
has a long lifetime and if, the neutral $B$ mixing is large. In 1987,
large $B^0-\bar{B}^0$ mixing was indeed observed by ARGUS at DESY. In
1999, the B factories started operation with the detectors, Babar at
SLAC and Belle at KEK. In the same year, direct CPV was clearly
confirmed in Kaon decays. The year 2001 saw the observation of large
CPV using the golden mode $B\to J/\psi K_{\! s}$ at both Belle and
Babar. This provided the first and unambiguous verification of the
complex phase in the KM proposal. In the years that followed, results
from Belle and Babar provided additional supporting evidence for the
KM scheme.
\section{CP Violation in the Standard Model}  
In the Standard model(SM), CPV arises from the complex Yukawa
couplings. The Yukawa interactions generate the mass terms. The
product of unitary matrices that diagonalize the mass matrices is the
Cabibbo-Kobayashi-Maskawa (CKM) matrix and it relates the weak
eigenstates to the mass eigenstates,
\[
{\small
\left( \begin{array}{l}
         d' \\ s' \\ b'
        \end{array} \right ) = V_{\rm CKM}\left( \begin{array}{r} d \\  s
          \\ b \end{array} \right ) = \left( \begin{array}{ccc}
           V_{ud} & V_{us} & V_{ub} \\
           V_{cd} & V_{cs} & V_{cb} \\
           V_{td} & V_{ts} & V_{tb} \\ \end{array} \right )
          \left( \begin{array}{r} d \\  s \\ b \end{array} \right )~.} 
\]
\label{eq:ckm}
~The elements of the CKM matrix describe the charged current couplings.
The matrix has to be unitary by construction. The orthonormality
relation obtained by using the first and third columns of $V_{\rm CKM}$ is
a simple complex relation,
$V_{ud}V_{ub}^*+V_{cd}V_{cb}^*+V_{td}V_{tb}^* =0$, which can be
geometrically represented by a triangle on a plane and is referred to as
the Unitarity triangle. This triangle is of relevance in $B$ decays
and all its sides are of the same order, resulting in large interior angles
$\alpha$, $\beta$ and $\gamma$, defined as,
\[
{\small
\alpha={\rm arg}
\left(\frac{-V_{tb}^*V_{td}}{V_{ub}^*V_{ud}} \right),~~\beta={\rm arg}
\left(\frac{-V_{cb}^*V_{cd}}{V_{tb}^*V_{td}} \right)~ {\rm and}~~
\gamma={\rm arg}
\left(\frac{-V_{ub}^*V_{ud}}{V_{cb}^*V_{cd}} \right)~.}
\]
The CKM has a hierarchical structure, in the Wolfenstein
representation it has the form\footnote{A few years ago, writing the elements of the matrix to $\cal{O}(\lambda^{\rm{3}})$ would have been sufficient,
but with increasing experimental precision and moreover since New
Physics effects are expected to be tiny, we write the elements upto
$\cal{O}(\lambda^{\rm{5}})$.}
\begin{eqnarray}
{\small
V_{\rm CKM}= \left( \begin{array}{ccc}
           1-\frac{\lambda^2}{2}-\frac{\lambda^4}{8} & \lambda & A\lambda^3(\rho-i\eta)  \\
           -\lambda+\frac{A^2\lambda^5}{2}[1-2(\rho+i\eta)] & 1-\frac{\lambda^2}{2}-\frac{\lambda^4}{8}(1+4A^2) & A\lambda^2 \\
            A\lambda^3[1-(1-\frac{\lambda^2}{2})(\rho+i\eta)]&
            -A\lambda^2+\frac{A\lambda^4}{2}[1-2(\rho+i\eta)]  &
            1-\frac{A^2\lambda^4}{2}\\ \end{array} \right ) +
        \cal{O}(\lambda^{\rm{6}})\nonumber}
\end{eqnarray}
Nonzero values of the phases ($\eta\neq 0$) imply CPV.  A huge effort
has been made at the $B$ factories and now also at the Tevatron to
measure the weak phases.

\section{Why and how to measure the phases?}
The KM mechanism for CPV is unique and predictive. Moreover, the CP
phases can be measured through certain asymmetries which are free of
hadronic uncertainties. Any inconsistencies if noted, would indicate
physics beyond the SM or New Physics (NP). Since the baryon number
density predicted by the KM mechanism is many orders of magnitude
below the observed value, we need new sources of CPV.  Correlations
among the many CPV observables in meson decays can possibly pinpoint
the kind of NP or at least constrain its parameters. Phases can be
observed only through interference terms in modes to which two (or
more) different amplitudes with distinct phases contribute.  The
different ways in which the interference terms appear, result in the
following categories of CPV:
\begin{enumerate}
\item CPV in Mixing: If the mass eigenstates differ from the
  CP eigenstates, it leads to a relative phase between the dispersive
  and absorptive parts of the transition amplitude from the neutral
  meson, $M^0$ to its conjugate 
  $\bar{M}^0$.
For the neutral $B$ it can be observed by measuring the semileptonic asymmetry,
\[
{\small
{\cal A}^{(q)}_{\mbox{{\scriptsize SL}}}\equiv
\frac{\Gamma(B^0_q(t)\to l^-\overline{\nu}_l X)-\Gamma(\overline{B^0_q}(t)\to
l^+\nu_l X)}{\Gamma(B^0_q(t)\to l^-\overline{\nu}_l X)+
\Gamma(\overline{B^0_q}(t)\to l^+\nu_l X)}=
\frac{|q/p|^4-1}{|q/p|^4+1},}
\]
which will be nonzero if $|q/p|\neq 1$. Current measurements imply
that ${\cal A}^{(d)}_{\mbox{{\scriptsize SL}}}$ is compatible with zero. 
In the K system, CPV in mixing had been seen,
\[
{\small
\!
\frac{\Gamma(K^0_L\to \pi^-l^+\nu_l)\!-\!\Gamma(K^0_L\to
\pi^+l^-\overline{\nu_l} X)}{\Gamma(K^0_L\to  \pi^-l^+{\nu}_l )\!+\!
\Gamma(K^0_L\to \pi^+ l^-\overline{\nu_l} )}\!=\!
\frac{1\!-\!|q/p|^2}{1\!+\!|q/p|^2}\!=\!(3.32\pm
0.06)\!\times\!10^{-3},}
\]
the measured value~\cite{PDG} is a weighted average of muon and electron measurements. 
\item CPV in Decay (Direct): For a decay amplitude with two weak
  contributions, the amplitude and its conjugate have the
  form\footnote{$\phi_{1,2}$ and $\delta_{1,2}$ are the weak and
    strong phases respectively.}, $A = A_1 e^{i \phi_1} e^{i \delta_1}
  +A_2 e^{i \phi_2} e^{i \delta_2}$, $ \bar{A} = A_1 e^{-i \phi_1}
  e^{i \delta_1} +A_2 e^{-i \phi_2} e^{i \delta_2}$.  The CP asymmetry
  is hence given by,
\begin{eqnarray}
A_{CP}=\frac{|A|^2-|\bar{A}|^2}{|A|^2+|\bar{A}|^2}
 = \frac{-2 A_1 A_2\sin(\Delta\phi)\sin(\Delta\delta)}{A_1^2+A_2^2+2A_1A_2\cos(\Delta\phi)\cos(\Delta\delta)\nonumber}~.
\end{eqnarray}
The asymmetry is non vanishing if $\Delta\phi=\phi_2-\phi_1=0$ and
$\Delta\delta=\delta_2-\delta_1=0$.  Direct CPV was clearly
established in $K$ decays with the accurate measurement of
$(\epsilon^\prime/\epsilon)$ in the year 1999. In many B decay modes
it has been measured to be significantly different from
zero~\cite{HFAG}. 
\item CPV due to interference between decays with and without mixing: 
Final states f into which both $M^0$ and $\bar{M}^0$ can decay, have
two interfering paths provided by the direct decay of $M^0\to f$ and
that of $M^0\to \bar{M}^0 \to f$.
The time dependent decay rate of $B^0\to f$ thus has the form,
\begin{eqnarray*}
\!\!\!\!\!\!\!\Gamma(B^0(t)\!\to\! f)= e^{-\Gamma t}\!\left[
\frac{|A|^2\!+\!|\bar{A}|^2}{2}\!-\!\frac{|\bar{A}|^2\!-\!|A|^2}{2}\!\cos(\Delta M
t)\!-\!{\rm Im}\Bigl(\frac{q}{p}\frac{\bar{A}}{A}\Bigr)\!\sin(\Delta M t) \right]~,
\end{eqnarray*}
while that for $B_s$ and $D$ decays is more complex, due to the width
difference $\Delta\Gamma\neq 0$. In the golden mode $B\to \psi K_
  s$, the decay amplitude is dominated by only one tree amplitude
with a real CKM element, implying that $\bar{A}=A$, the time dependent
CP asymmetry has the simple form, $a_{CP}=\sin(2\beta)\sin(\Delta M
t)$ and is almost completely free of hadronic uncertainties (measures
$\beta$ cleanly up to $\approx 1\%$).  Time dependent CP asymmetery has
been measured in many other $B$ decay modes, $\pi\pi$, $\phi K_{\! s}$ etc.
\end{enumerate}
\section{Current Status and what it implies}
\begin{figure}[htbp]
\begin{minipage}{0.59\linewidth}
\centering
\includegraphics[scale=0.19]{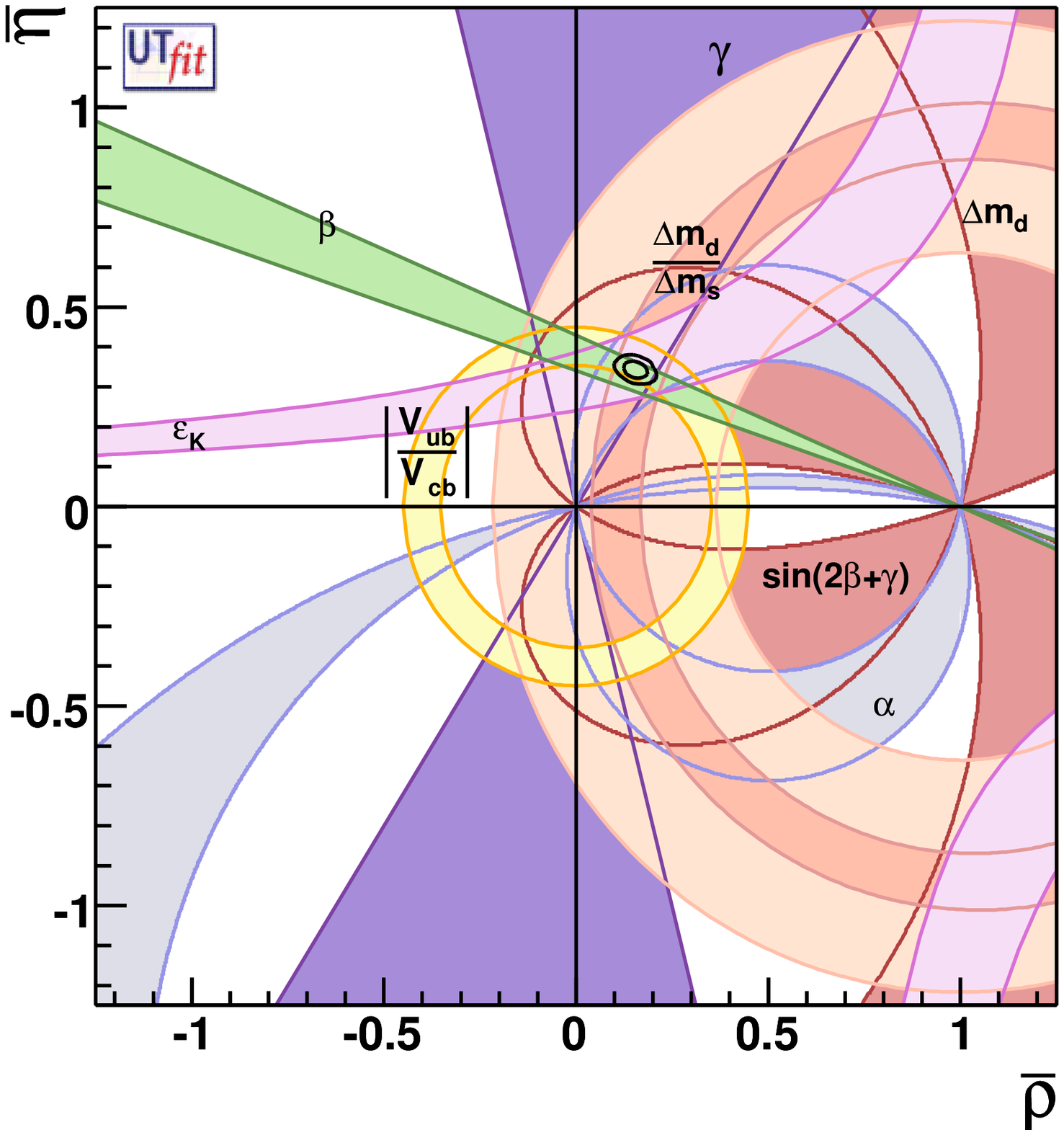}
\includegraphics[scale=0.19]{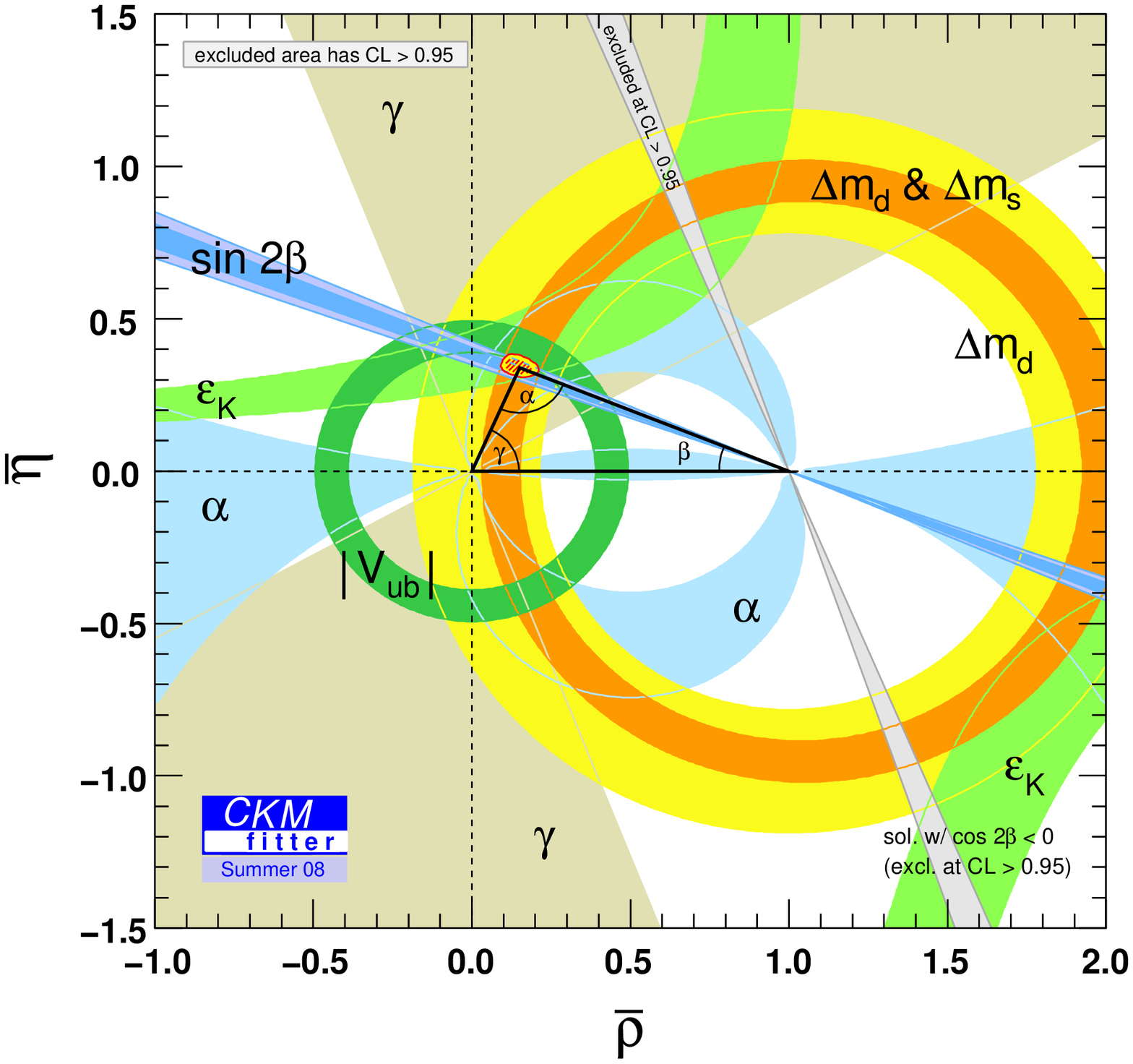}
\caption{Constraints in the $\bar{\rho}-\bar{\eta}$ plane from global
  fits from Utfit(left) and CKMfitter(right)}
\label{fig:global}
\end{minipage}%
\hskip 0.4cm
\begin{minipage}{0.38\linewidth}
\centering
\includegraphics[scale=0.21]{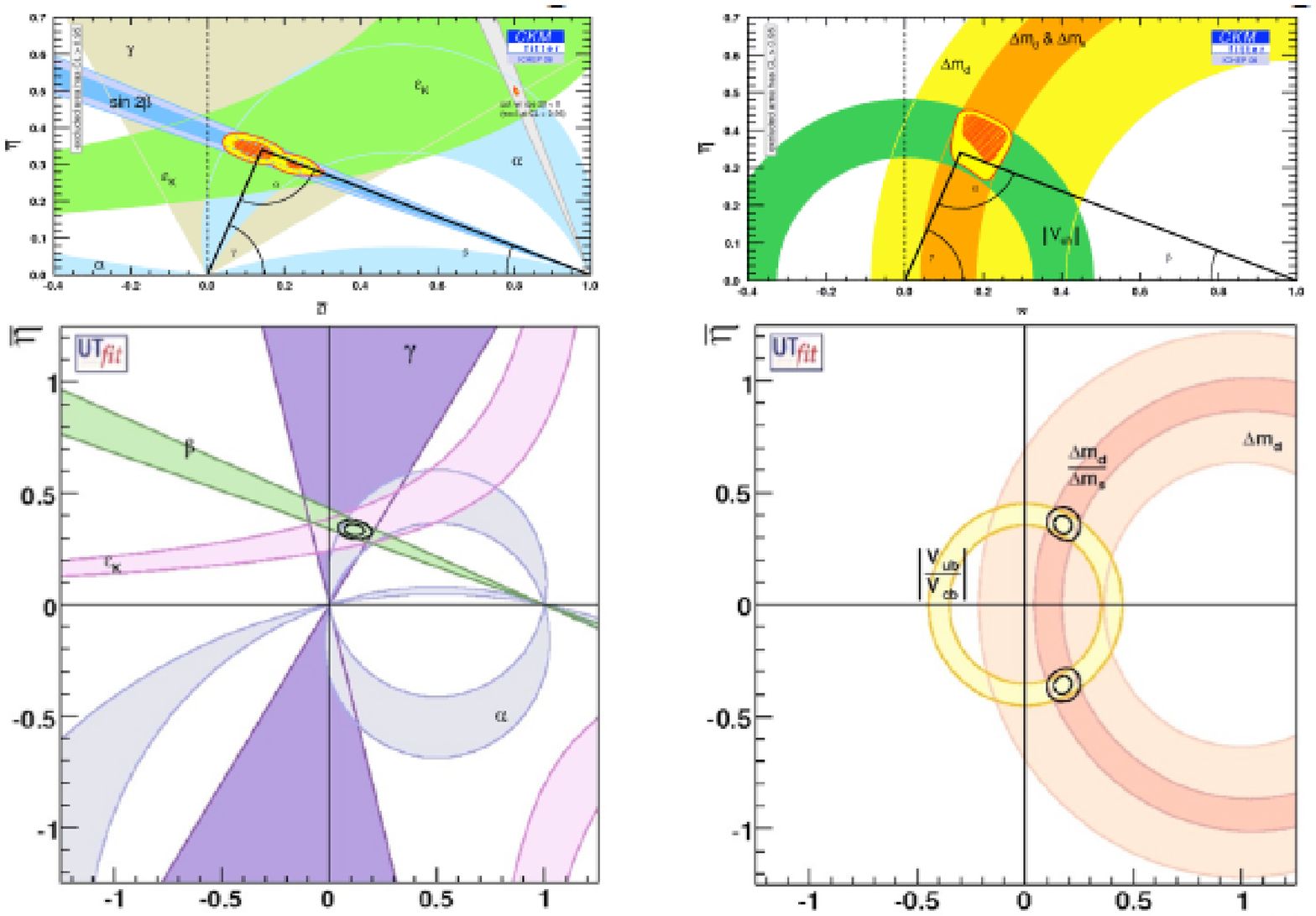}
\caption{Constraints from CKMfitter(up) and Utfit(down) in $\bar{\rho}-\bar{\eta}$ plane using CP violating(left) and CP conserving(right) observables.}
\label{fig:CP}
\end{minipage}
\end{figure}
Current data from $K$ and $B$ decays have been used by the UTfit
collaboration~\cite{utfit} and the CKMfitter group~\cite{ckm}
to perform a global fit in the $\bar{\rho}-\bar{\eta}$ plane,
depicting the allowed region for the apex of the unitarity triangle.
The resulting plots are shown in Fig.~\ref{fig:global} and clearly
show that all observations are consistent with each other.
Fig.~\ref{fig:CP} shows the contributions from CP conserving and CP
violating observables, again these are not only individually
consistent but also with each other. Current observations hence imply
that:
\begin{itemize} 
\item [$\bullet$] $\eta$ is non vanishing, implying that the KM mechanism is
  working.
\item [$\bullet$] There is consistency of results from K mesons and B mesons.
\item [$\bullet$] Almost all CP violating observables (tree level and
  loop level) as well as measurements of rare decays and mixing
  parameters, are consistent with the KM mechanism. The CKM picture of
  SM seems to have been successful!
\item [$\bullet$] CKM mechanism is the dominant mechanism for CP violation and flavor mixing.
\item [$\bullet$] BUT, there is room for New Physics.
\end{itemize}
\section{Hints of New Physics?}
\begin{enumerate}
\item The $\Delta A_{K\pi}$ puzzle: The CP asymmetry for $B^0\to
  K^+\pi^-$ and for $B^+\to K^+\pi^0$ is expected to be the same from
  isospin. However, the measured asymmetries~\cite{paoti} for these
  modes are: $A_{CP}(K^+\pi^-)= -0.098^{+0.012}_{-0.011}$ and
  $A_{CP}(K^+\pi^0)= 0.050\pm0.025$. The difference in these
  asymmetries, $\Delta A_{K\pi} = -0.147\pm 0.028$, is non vanishing at
  $5.3\sigma$. Note that the $B^0\to K^+\pi^-$ mode gets contributions
  from Penguin and Tree diagrams, while the $B^+\to K^+\pi^0$ mode can
  have additional colour suppressed and Electroweak
  penguin contributions (ignoring smaller contributions).
  There have been attempts to explain the discrepancy, through NP
  using fourth generation~\cite{4gen}.  More recently~\cite{Ciuchini},
  it has been indicated that the theory calculations are consistent,
  if sizable $\Lambda/m_b$ corrections are taken into account. Since,
  hadronic uncertainties are involved, predictions for CP violating
  asymmetries have large errors, so a firm conclusion cannot yet be
  made.
\item New Physics in $\Delta S$: For a final state $f$, the deviation
  of the time dependent CP asymmetry for $f$ from that of $\psi
  K_{\! s}$ mode, $\Delta S_f = -\eta_f S_f - S_{\psi K_{\! s}}$
  is expected to be zero. While many modes are now consistent with
  this expectation, there are slight deviations in some modes. While
  theory predictions are for small and positive deviations,
  experimental results yield negative deviations in some
  modes~\cite{Beneke}. NP scenarios have been proposed to explain this
  discrepancy~\cite{np}.  Although the effects now seem to be tiny,
  with small NP effects expected, we need to wait for improved
  statistical significance. In addition, there is a discrepancy in
  $S_{\psi K_{\! s}}$ versus $\beta$ measured from tree level
  measurements alone~\cite{betatree}.
\item $\beta_s$ measurement at the Tevatron: CDF and D0 have performed
  a time dependent analysis of $B_s\to \psi\phi$, to get a correlated
  measurement of ($\beta_s$, $\Delta\Gamma$). The Utfit combination of
  the Tevatron data gives a $2.9\sigma$ deviation~\cite{tarantino} of
  $\phi_{\beta_s}$ from SM.
\end{enumerate} 
\section{What kind of New Physics is possible?}
In the Standard Model, there is no guiding flavour principle. The
pattern of masses and mixing parameters is unexplained. It is unclear
why there should be only three generations. The current baryon number
density, $n_B/n_\gamma = (5.5\pm 0.5)\times 10^{-10}$, reflects the
baryon asymmetry induced by baryogenesis. New sources of CPV
are required to generate this observed baryon asymmetry.  Apart from
these questions in the flavour sector, there are many other reasons
which lead us to believe that the SM is a low energy effective theory
of some fundamental theory. In particular, quadratic divergence in the
SM Higgs mass, requires NP at around the TeV scale.

Hence, we can extend the SM Lagrangian by higher dimension operators,
suppressed by powers of the NP scale. Most rare decay modes are
flavour changing neutral current (FCNC) processes that appear only at
the loop level and are hence useful for NP searches. For example,
consider the $\Delta F =2$ processes of mixing involving the down type
quark (D). In the SM, the Lagrangian has the form, $-{\cal L}_{eff} =
\frac{C_0}{4 \Lambda_0^2}(V_{ti}^*V_{tj})[\bar{D}_{Li}\gamma_\mu
D_{Lj}]^2$, where, $C_0\sim\cal{O(\rm 1)}$ and $\Lambda_0$ is the
scale for loop suppressed SM processes. Assuming that the NP effective
operator has the same Dirac structure as in the SM, we have $-{\cal
  L}_{eff}^{NP} = \frac{C_{NP}}{\Lambda_{NP}^2}[\bar{D}_{Li}\gamma_\mu
D_{Lj}]^2$ with coefficient, $C_{NP}\sim\cal{O(\rm 1)}$ and
$\Lambda_{NP}$ of the order of the mass of the NP particle. The
various measurements ($\Delta m_{B_s}$, $\Delta m_{B}$, $\Delta
m_{D}$, $\Delta m_{K}$, $\epsilon_K$) of the $\Delta F =2$ processes,
imply that the NP scale must be above $10^2 - 10^4$ TeV, much larger
than the weak scale. Thus, NP with generic flavour violation structure
is excluded at the TeV scale. This is the New Physics Flavour
puzzle~\cite{MFV}.  

It can be resolved by having Minimal Flavour violation(MFV). If the
scale of NP has to be of TeV order, we need some principle to make the
coefficients of FCNC's small. In the SM, the global flavor symmetry
group, $G_F = U(3)_Q\times U(3)_U\times U(3)_D\times U(3)_L\times
U(3)_E$ is broken only by Yukawa couplings $Y_U$ and $Y_D$.  In the
MFV hypothesis, there is a unique source of breaking of $G_F$,
operators that break $G_F$, must transform just as the Yukawa terms.
It was formalized by D'Ambrosio et al~\cite{ambrosio}, who suggested
that the Yukawa couplings be promoted to spurions that transform under
$G_F$ as, $Y_U\!  \sim\!(3, \bar{3}, 1), Y_D\!\sim\!(3, 1, \bar{3})$
(for the quark sector). MFV NP is also formally invariant under $G_F$,
breaking coming only from insertions of spurion fields $Y_{U,D}$.
Integrating out heavy fields (NP fields, Higgs, top, W and Z) leads to
a low energy effective field theory invariant under $G_F$.  Using the
basis in which $Y_D\!=\!\lambda_D$ and $Y_U\!=\!V^\dagger\lambda_U$,
where $\lambda_D$ and $\lambda_U$ are diagonal matrices proportional
to quark masses and $V$ is the CKM matrix, insertions of $(Y_U
Y_U^\dagger)_{ij}$ will be of the order $\lambda_t^2 V_{ti}^*V_{tj}$,
making this theory very predictive.  In addition, if one imposes the
constraint that the structure of low energy operators be the same as
in SM and only the Wilson coefficients of weak operators deviate from
SM values, one gets the constrained MFV.  This is clearly
experimentally distinguishable, due to correlations between
observables.
\section{How do we look for New Physics?}
In order to find NP, we need to make even more precise measurements of
the CP violating phases. Measurement of Rare decays : $B_s\to
\mu^+\mu^-$, $B\to\tau\nu$, $B\to K^*\ell^+\ell^-$, $B\to
X_{s,d}\nu\bar{\nu}$, $B\to X_{s,d}\gamma$, $K\to\pi\nu\bar{\nu}$,
$\mu\to e\gamma$ etc., have played an important role in constraining
NP models and further precision measurements could pinpoint to NP (MFV
or nonMFV) or at least narrow down the parameter space of various NP
models. As an example, apart from the forward backward asymmetry being
currently measured in $B\to K^*\ell^+\ell^-$~\cite{paoti}, a detailed
angular analysis of this mode can be used to determine various other
asymmetries~\cite{KSSS}. Some of these asymmetries are CP violating
and are expected to be negligible in the SM, a measured value would
imply NP.  In the $D$ system, within the SM CPV is negligible and an
observation of CPV would be a clear signal of NP~\cite{CPVD}. A technique to
accurately determine all the mixing parameters, including the CP
violating phase has been given in Ref.~\cite{SSBPD}. With higher
statistics possible at LHCb and Super B factories, all such searches
for NP should be feasible!
\section{Conclusions}
The Kobayashi Maskawa mechanism of CP violation has been well tested
with the results from the B factories. However, to explain the baryon
asymmetry, we expect new sources of CP violation. Search for new
physics, requires precision measurements of the CP violating
parameters. In conjunction with rare B and K decays, it could point to
the kind of new physics present: SUSY, Extra Dimensions, Little Higgs
$\ldots$. We look forward to data from LHCb and Super B factories to
achieve this goal.
%
%
%

%
\end{document}